\documentclass[global, twocolumn]{svjour}

\usepackage{graphicx}

\begin{document}

\title{An optical conveyor belt for single neutral atoms}
\titlerunning{An optical conveyor belt for single neutral atoms}

\author{Dominik Schrader, Stefan Kuhr, Wolfgang Alt, Martin M\"uller, Victor Gomer, and
Dieter Meschede} \institute{
 Institut f\"ur Angewandte Physik, Universit\"at Bonn\\
 Wegelerstr. 8, D-53115 Bonn, Germany\\
  \email{schrader@iap.uni-bonn.de}
}

\authorrunning{D. Schrader {\it et al.}}
\date{July 5, 2001}

\maketitle

\begin{abstract}
Using optical dipole forces we have realized controlled transport
of a single or any desired small number of neutral atoms over a
distance of a centimeter with sub-micrometer precision. A standing
wave dipole trap is loaded with a prescribed number of cesium
atoms from a magneto-optical trap. Mutual detuning of the
counter-propagating laser beams moves the interference pattern,
allowing us to accelerate and stop the atoms at preselected points
along the standing wave. The transportation efficiency is close to
100~\%.  This optical "single-atom conveyor belt" represents a
versatile tool for future experiments requiring deterministic
delivery of a prescribed number of atoms on demand.
\end{abstract}


%\section{Introduction}

Quantum engineering of microscopic systems requires manipulation
of all degrees of freedom of isolated atomic particles. The most
advanced experiments are implemented with trapped chains of ions
\cite{Neuhauser 80,Blatt 99,Sackett 00}. Neutral atoms are more
difficult to control because of the weaker interaction of induced
electric or paramagnetic dipoles with inhomogeneous
electromagnetic fields. Optical dipole traps \cite{Chu 86} could
provide a level of control similar to ion traps, since they store
neutral atoms in a nearly conservative potential with long
coherence times \cite{Davidson 95}. The variety of different
dipole traps allows for an individual design, depending on the
specific experimental demands \cite{Grimm 00}.

Here we use a time-varying standing wave optical dipole trap to
displace atoms by macroscopic distances on the order of a
centimeter with sub-micrometer precision \cite{Kuhr 01}. A similar
technique of moving optical lattices has been applied for the
acceleration of large ensembles of atoms in \cite{Dahan
96,Wilkinson 96}. Time-dependent magnetic fields can also be used
for controlled transport of clouds of neutral atoms as has been
demonstrated with a micro-fabricated device \cite{Haensel 01}.
Recently, techniques have been developed to load a dipole trap
with single atoms only \cite{Frese 00,Schlosser 01}. In our case,
we have combined controlled manipulation of the trapping potential
with deterministic loading of a dipole trap with a prescribed
small number of atoms \cite{Kuhr 01}. These atoms are then
transported with high efficiency over macroscopic distances and
observed by position-sensitive fluorescence detection in the
dipole trap. Here we describe the transportation technique in
detail. We also analyze the dependency of the measured
transportation efficiency on both the transportation distance and
the acceleration. This is followed by a brief discussion of
possible applications and alternative approaches.

\section{The standing-wave dipole trap}

\begin{figure}[!h]
 \centering
\includegraphics{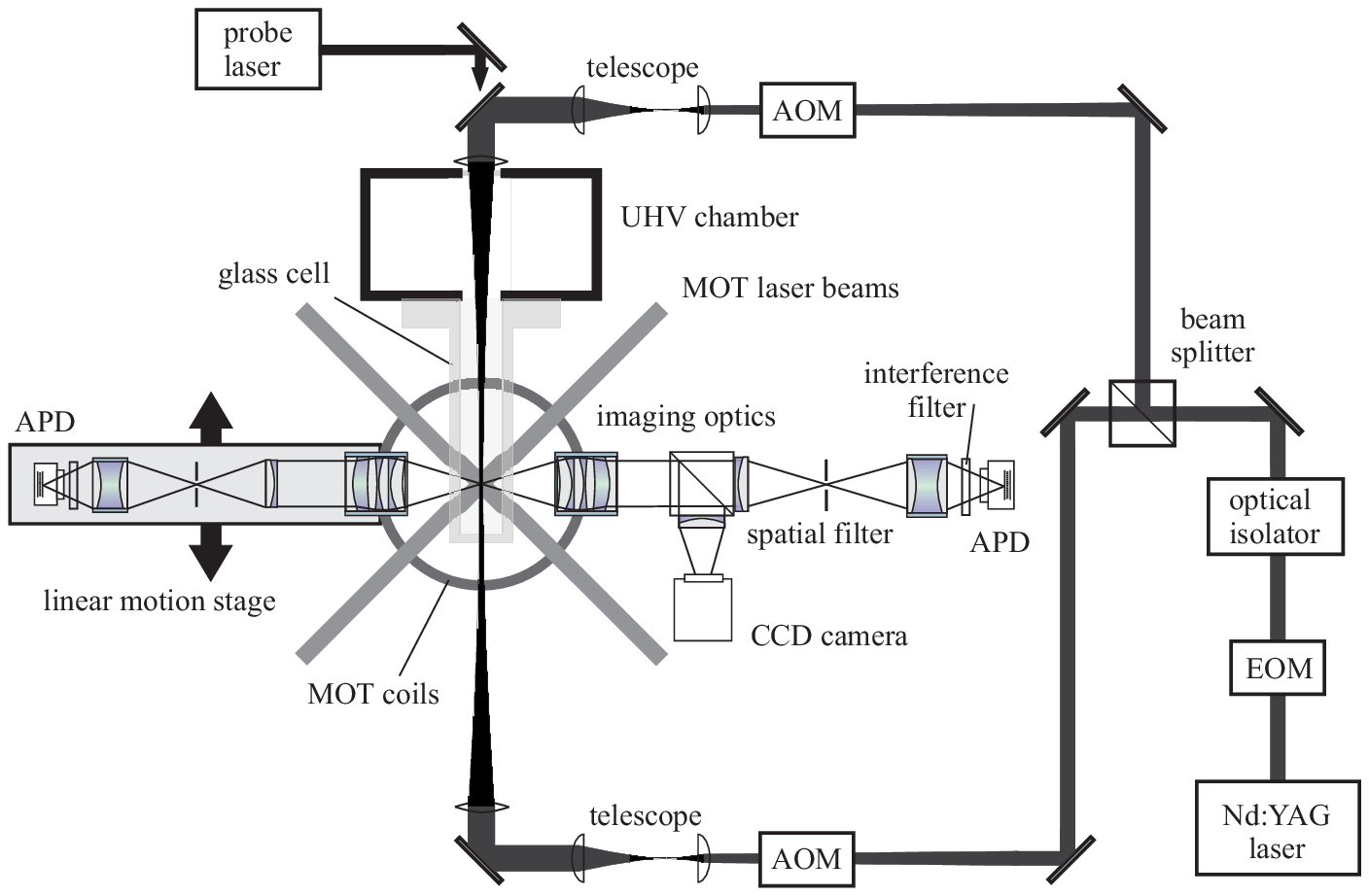}
\caption{Scheme of the experimental setup. The fixed imaging
optics on the right side is used to monitor the fluorescence at
the MOT position. The imaging optics mounted on the linear motion
stage on the left side is used for atom detection at any spot
along the dipole trap.}\label{figure1}
\end{figure}

Our dipole trap (Fig.\ref{figure1}) consists of two
counter-propa\-gating Gaussian laser beams with equal intensities
and optical frequencies $\nu_1$ and $\nu_2$ producing a position-
and time-dependent dipole potential
\begin{equation}
\label{dipole-potential}
 U(\rho,z,t) = U_0 \frac{ w_0^2}{w(z)^2}\ e^{-
\frac{2\rho^2}{w(z)^2} } \cos^2(\pi \Delta \nu t - kz).
\end{equation}
The optical wavelength is $\lambda=2\pi/k$, $w^2(z)=w_0^2 \,
(1+z^2/z_0^2)$ is the beam radius with waist $w_0$ and the
Rayleigh length $z_0 = \pi w_0^2/\lambda$, and $\Delta \nu =\nu_1
- \nu_2 \ll \nu_1, \nu_2$ is the mutual detuning of the laser
beams. The laser beams have parallel linear polarization and thus
produce a standing wave interference pattern. Changing the
frequency difference $\Delta \nu$ moves the stationary ($\Delta
\nu$=0) standing wave along the $z$-axis. This can be understood
intuitively in a simple picture. Assume, the two beams are detuned
by $-\Delta\nu/2$ and $+\Delta\nu/2$, respectively. In a reference
frame moving with velocity $v=\lambda~\Delta\nu/2$ both beams are
Doppler shifted by the same amount resulting in a stationary
standing wave \cite{Wilkinson 96}. In the laboratory frame, this
corresponds to a motion of the standing wave along the optical
axis with velocity $v$. In our experimental realization, as
described below, it is more convenient to detune only one of the
beams by $\Delta\nu$ while keeping the other one at constant
frequency.

Both dipole trap laser beams are derived from a single Nd:YAG
laser ($\lambda=1064$~nm) which is far red detuned from the
$6S_{1/2}\rightarrow6P_{1/2,3/2}$ transitions of cesium
($\lambda_{\rm D1}=894$~nm, $\lambda_{\rm D2}=852$~nm). Since the
laser detuning is much larger than the atomic fine-structure
splitting, the maximum potential depth $U_0$ is
\begin{equation}
 U_0 = \frac{\hbar \Gamma}{2} \frac{P}{\pi w_0^2 I_0}
 \frac{\Gamma}{\Delta}.
\end{equation}
Here, $\Gamma = 2\pi\cdot 5.2$~MHz is the natural linewidth of the
cesium D$_2$-line and $I_0 =1.1$~mW/cm$^2$ is the corresponding
saturation intensity. The effective laser detuning $\Delta$ is
given for alkalis by \cite{Grimm 00}
$\Delta^{-1}=(\Delta_1^{-1}+2\Delta_2^{-1})/3$, where $\Delta_{\rm
i}$ is the detuning from the D$_{\rm i}$-line. In our case $\Delta
\approx 10^7 \Gamma$. For a total power $P$ of 4~W and a beam
waist $w_0$ of 30~$\mu$m, the potential depth $U_0$ is 1.3~mK.

The maximum photon scattering rate $\Gamma_{\rm sc}$ is
proportional to the potential depth
\begin{equation}
  \Gamma_{\rm sc}=\frac{\Gamma}{\Delta}\frac{U_0}{\hbar}
\end{equation}
and amounts to 15~photons/s for our parameters. The spin
relaxation rates, however, are two orders of magnitude smaller
than $\Gamma_{\rm sc}$. In previous experiments in a travelling
dipole trap we measured spin relaxation times on the order of
several seconds \cite{Frese 00}.

In a harmonic approximation an atom of mass $m$ (for a cesium atom
$m=2.2 \cdot 10^{-25}$~kg) trapped in such a standing wave
oscillates with frequencies $\Omega_{\rm
z}=2\pi\sqrt{2U_0/m\lambda^2} \approx 2\pi \cdot 340$ kHz in axial
and $\Omega_{\rm rad}=$

\noindent
$\sqrt{4U_0/m w_0^2} \approx 2\pi \cdot 2.6$~kHz in
radial directions. The rms size of the ground state wavefunction
in the axial  and radial directions is $\Delta z_0 =
\sqrt{\hbar/2m\Omega_{\rm z}}$ = 11~nm and $\Delta \rho_0=120$~nm,
respectively. In our case, atoms at about Doppler temperature
($T_{\rm D}=\hbar\Gamma/2 k_B=125$~$\mu$K) \cite{Alt 01b} are
localized in the axial direction to $43$~nm and in the radial
direction to $5.6$~$\mu$m.

The experimental setup is shown in Fig. 1. Acousto-optical
modulators (AOMs) control the frequencies of both laser beams that
generate the dipole trap. They are driven by a digital
dual-frequency synthesizer (by APE Berlin) with two
phase-synchronized RF-outputs. Both AOMs are set up in double-pass
configuration in order to avoid beam walk-off during a frequency
sweep. To achieve optimal interference contrast of the standing
wave, the optical path lengths of both laser beams are equalized.
A coherence length of the Nd:YAG laser of $>10$~cm ensures a
fringe visibility close to 100~\% within the desired displacement
distance.

A standard six-beam magneto-optical trap (MOT) \cite{Raab 87} at
the center of a UHV glass cell serves as our primary source of
single cold atoms \cite{Frese 00}. Dissipative forces slow cesium
atoms from the background vapor and cool them down to about
Doppler temperature. The high magnetic field gradient of
400~Gauss/cm localizes the trapped atoms to a region of diameter
30~$\mu$m which is much smaller than the 2~mm waist of the MOT
lasers. It also provides a low loading rate of 2~atoms/min only
which ensures that accidental loading events during the
measurement procedure are negligible. We speed up the loading
process by temporarily lowering the magnetic field gradient to
40~Gauss/cm. This results in a larger capture cross section which
significantly increases the loading rate. After several 100~ms the
field gradient is returned to its initial value, concentrating the
trapped atoms at the center of the MOT. Varying the time during
which the field gradient is low enables us to select a specific
{\it mean} atom number. This procedure allows us to repeat the
experiment with identical parameters in quick succession many
hundreds of times. Since we record the initial number of atoms
trapped in the MOT, we can evaluate the results for each atom
number separately. By slightly increasing the cesium partial
pressure in the vacuum chamber this system could easily deliver
one atom within 100~ms.

The fluorescence light of the atoms is collected by a diffraction
limited objective \cite{Alt 01} (NA=0.29) and projected onto an
avalanche photodiode (APD, model SPCM-200 by EG\&G) with a quantum
efficiency of 50~$\%$ at $\lambda=852$~nm. This yields a photon
count rate of $5\cdot10^4s^{-1}$ per atom. Spatial filtering
reduces the MOT laser stray light background to a count rate of
$2\cdot 10^4s^{-1}$, allowing us to determine the exact number of
trapped atoms in real time, with a typical uncertainty of
$<1$~$\%$ in 1~ms. Interference filters transmitting the
fluorescence light at 852~nm attenuate the strong Nd:YAG laser
stray light to 30~photons/s which is as low as the dark count
rates of the APDs.

\begin{figure}
 \centering
\includegraphics{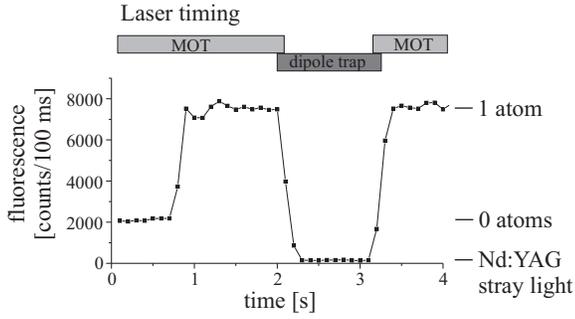}
\caption{Storage of a single atom in the dipole trap for 1~s. The
fluorescence signal of the atom demonstrates the trapping in the
dipole trap and recapturing by the MOT.}\label{figure2}
\end{figure}

Transfer of atoms from the MOT into the optical dipole trap with a
high efficiency is the backbone of the experiment. A prerequisite
for efficient transfer is a thorough alignment of the dipole trap
laser onto the MOT. As a sensitive alignment criterion we use the
fact that the Nd:YAG laser shifts the atomic transition out of
resonance, which lowers the fluorescence rate of the MOT. The
dipole trap laser is therefore superposed with the MOT by
minimizing the fluorescence rate of a single trapped atom
\cite{Frese 00}. This is done for both dipole trap laser beams
separately. Since the localization of the atom in the MOT is
tighter than the foci of the two beams, this alignment also yields
their optimal mutual superposition to provide the standing wave
structure. To transfer cold atoms from the MOT into the dipole
trap, both traps are simultaneously operated for several
milliseconds before we switch off the MOT (Fig.~\ref{figure2}).
After storage in the dipole trap the atoms are transferred back
into the MOT by the reverse procedure. In other experiments only a
fraction of atoms stored in a MOT can be loaded into a dipole trap
due to inelastic collisions, see \cite{Kuppens 00} and references
therein. In our case, however, the small number of atoms together
with the perfect superposition of the two small traps and
intrinsic cooling during the transfer process \cite{Frese 00}
warrant a transfer efficiency of nearly 100~$\%$.

The ability to transfer atoms between the two traps provides a
simple procedure for measuring the lifetime of the atoms in the
dipole trap. Background gas collisions limit this trap lifetime to
about 25~s. Other heating mechanisms such as photon scattering
(about 1.5~$\mu$K/s in our case), intensity fluctuations and beam
pointing instabilities of the trapping laser beams \cite{Gehm 98}
are not observable in our experiment. However, fluctuations of the
relative phase with an rms-value of roughly $2\pi/1000$ between
the two RF-outputs of the frequency synthesizer are directly
translated into position fluctuations of the dipole trap
potential. This causes heating of the trapped atoms \cite{Alt 01b}
and limits the lifetime to 3~s, which is still several orders of
magnitude longer than all experimentally relevant time scales.

\section{Transportation efficiency}

\begin{figure}
 \centering
\includegraphics{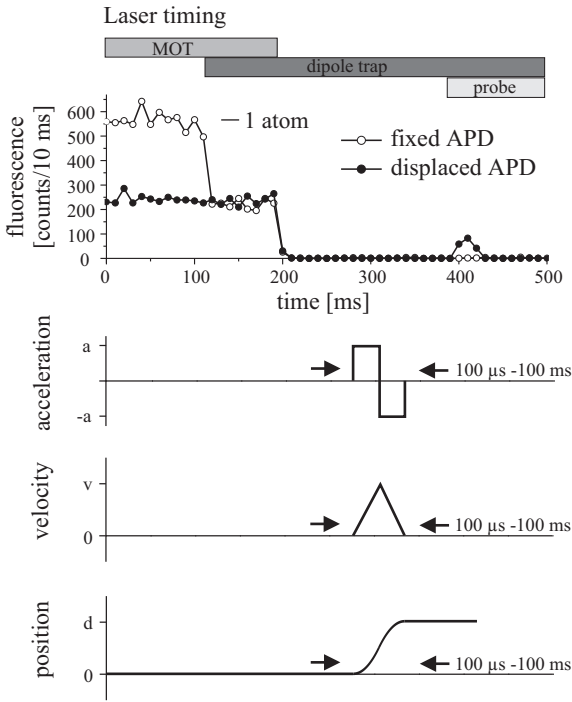}
\caption{Single-atom conveyor belt. The fluorescence signal of the
atom is recorded by both APDs during the transportation sequence.
The fixed APD initially confirms the presence of the atom in the
MOT. During transfer into the dipole trap the MOT fluorescence is
decreased due to the light shift. Initially, the displaced APD
(filled circles) does not see the trapped atom but only detects
stray light of the MOT laser beams. The burst of fluorescence at
$t=400$~ms originates from the same atom displaced by 1~mm, which
is illuminated in the dipole trap with a resonant probe
laser.}\label{figure3}
\end{figure}

The conveyor belt accelerates a trapped atom and brings it to a
stop at preselected points along the standing wave. For this
purpose, we transfer one atom from the MOT into the dipole trap
(Fig. \ref{figure3}) before the MOT lasers and magnetic field are
switched off. To move the atom over the distance $d$ it is
uniformly accelerated along the first half of $d$ and decelerated
in the same manner along the second half.

To accomplish this, the digital frequency synthesizer linearly
sweeps the frequency of one of the modulators in a
phase-continuous way from $f_0$ to $f_0 + \Delta f_{\rm max}$ and
back to $f_0$ (Fig.~\ref{figure3}), while the other modulator
remains at $f_0$. Since the AOMs are set up in double pass
configuration, the maximum relative detuning of the AOM
frequencies $\Delta f_{\rm max}$ is translated into an optical
detuning of $\Delta\nu_{\rm max}=2\Delta f_{\rm max}$. These
frequency sweeps accelerate and decelerate the standing wave
structure achieving a maximum velocity of $v=\lambda\Delta\nu_{\rm
max}/2$. The duration of the overall displacement procedure,
$t_{\rm d}$, determines the required accelerations $a=\pm
\lambda\Delta \nu_{\rm max}/t_{\rm d}$. The moving potential wells
of the dipole trap thus carry the atom along the required distance
$d=a t_{\rm d}^2 / 4$. This distance can be controlled with
sub-micrometer precision by heterodyning both frequencies of the
AOM drivers. A counter monitors the number of cycles during a
frequency sweep, which directly measures the transportation
distance in multiples of $\lambda$.

To observe the atom at its new position, we use a second optical
system identical to the one used for collecting fluorescence from
the MOT (Fig.~\ref{figure1}). Spatial filters limit the field of
view to a radius of $\sim40~\mu$m, which is much smaller than the
typical displacements. A linear motion stage moves both detector
and imaging optics by the transportation distance. The fixed
imaging optics permanently monitors the MOT region, both to verify
the initial presence of a single atom in the MOT and to confirm
its absence after displacement, see Fig.~\ref{figure3}. At its
destination, the transported atom is illuminated by a resonant
probe laser ($F=4\rightarrow F'=5$ transition of the D$_2$ line)
overlapped with a repumping laser ($F=3\rightarrow F'=4$),
providing cyclic optical excitation. Both probe and repumping
laser are collimated and overlapped with the Nd:YAG laser and
focused to a beam diameter of only 100~$\mu$m in order to achieve
a high intensity of $10\,I_0$ at the position of the atoms without
a measurable contribution to stray light. From a single atom we
routinely collect 40~fluorescence photons within 40~ms with near
zero background. This allows us to unambiguously detect the atom
at its new position as long as $d\leq 3$~mm, as will be shown
below.

\begin{figure}[!t]
 \begin{center}
\includegraphics{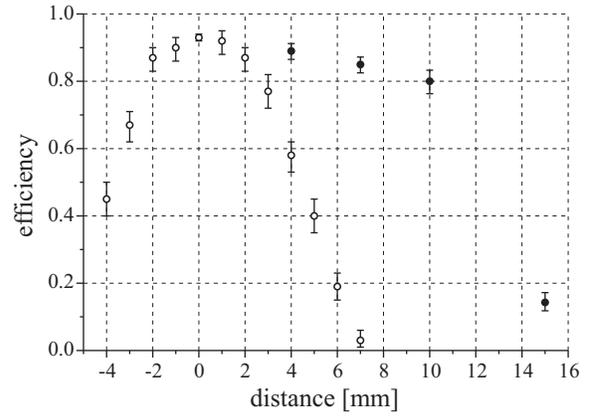}
\end{center}
\caption{Transportation efficiency of the optical con\-veyor-belt
for a constant acceleration of 500~m/s$^2$. Each data point
results from $\sim$100 shots performed with one atom each. Empty
circles: The atom is detected by resonant illumination at its new
position. Filled circles: More efficient detection by moving the
atom back and recapturing it into the MOT.}\label{figure4}
\end{figure}

This detection scheme demonstrates the deterministic delivery of a
single atom to a desired position. The measured probability to
observe the transported atom as a function of the displacement is
shown in Fig. \ref{figure4} as empty circles. For small distances,
the fraction of detected atoms is above 90~\%. However, the
position dependence of the trap depth limits this detection
efficiency for larger displacements from the laser focus. The
tight focusing of the trapping laser beams yields a Rayleigh
length $z_0$ of only 3~mm. Due to the divergence of the beams, the
local trap depth $U(z)$ scales with the displacement $z$ from the
focus as
\begin{equation}\label{localdepth}
   U(z)=U_{0}\left(1+\frac{z^2}{z_0^2}\right)^{-1}.
\end{equation}
During resonant excitation the atom is heated by scattering
photons. The fluorescence signal lasts until the atom is
evaporated out of the trap, which happens on average after $N=U(z)
/2E_{\rm r}$ scattering events. Here $E_{\rm r}=(\hbar k_{\rm
D2})^2/2m$ is the photon recoil energy with $k_{\rm
D2}=2\pi/\lambda_{\rm D2}$. As a consequence, we observe that the
number of detected fluorescence photons per atom is proportional
to $U(z)$. In this measurement we detect the presence of the atom
if a fluorescence peak substantially exceeds (more than 5 photons)
the stray light background (2 photons on average) of the Nd:YAG
and the probe laser beams. Thus, at $d > 3$~mm, the probe laser
can evaporate the atom out of the dipole trap before enough
fluorescence photons have been detected.

The actual transportation efficiency, however, is much higher than
that shown by the resonant illumination detection. To demonstrate
this, we use the MOT to detect the atom with 100~\% efficiency.
Without resonant illumination, the displaced atom is transported
back to $z=0$ before we switch on the MOT lasers to reveal the
presence or absence of the atom. The results of this measurement
are shown in Fig. \ref{figure4} as filled circles. Even for
distances as large as 10~mm, the two-way transportation efficiency
remains above 80~\%. At a distance of 15~mm, however, the
transportation efficiency drastically decreases to 16~$\%$.

\begin{figure}
 \begin{center}
\includegraphics{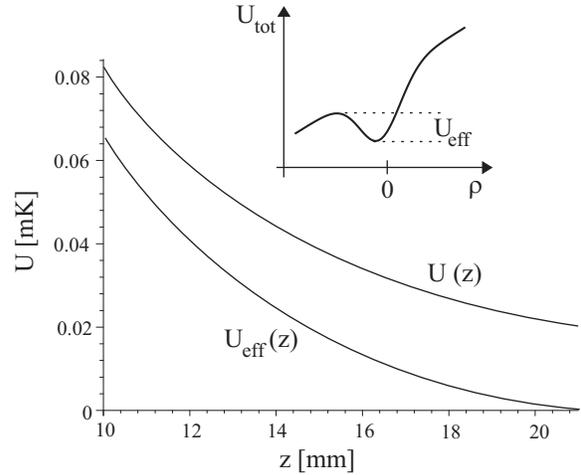}
\end{center}
\caption{Reduction of the trap depth due to gravity. The inset
shows the sum $U_{\rm tot}$ of the dipole potential and the
gravitational potential versus the radial coordinate $\rho$ for an
arbitrary potential well along the optical $z-$axis. Below, the
resulting effective potential depth $U_{\rm eff}(z)$ is compared
to the pure dipole potential $U(z)$ for $z=10-21$~mm. The position
$z=21$~mm, where $U_{\rm eff}$ completely disappears, represents
the fundamental upper bound for the transportation distance in our
dipole trap.}\label{figure5}
\end{figure}

The atoms are lost at this distance because gravity reduces the
effective potential depth. In our setup, the optical $z$-axis of
the dipole trap is oriented horizontally. Thus, the potential in
vertical direction is the sum of the radial dipole trapping
potential and the gravitational potential
\begin{equation}
 U_{\rm tot}(\rho,z)=U(z)
 e^{-\frac{2\rho^2}{w(z)^2}}+mg\rho,
\end{equation}
where $U(z)$ is the trap depth of a potential well, given in Eq.
(\ref{localdepth}). The acceleration due to gravity tilts these
Gaussian potential wells, which reduces the trap depth to $U_{\rm
eff}(z)$, see Fig. \ref{figure5}. In contrast to the pure
Lorentzian dipole potential $U(z)$, this effective potential
disappears at $z = 21$~mm. However, due to the initial energy of
the atom, we lose the atom at even smaller distances. This
interpretation is supported by an independent measurement
\cite{Alt 01b} in a stationary standing wave. We measured the
survival probability of the atom after an adiabatic lowering of
the trap depth  by attempting to recapture it in the MOT. We
observed that 80~$\%$ of the atoms survive if the trap depth is
reduced from $U_0$ to $0.03\,U_0$, which equals the effective
potential depth $U_{\rm eff}(z)$ at $z=13$~mm. However, a
reduction to $0.01\,U_0$, which corresponds to a displacement to
$z=17$~mm, yielded a survival probability of only 10~$\%$. This is
in agreement with the measured transportation efficiency of 16~\%
at $z=15$~mm.

\begin{figure}
 \centering
\includegraphics{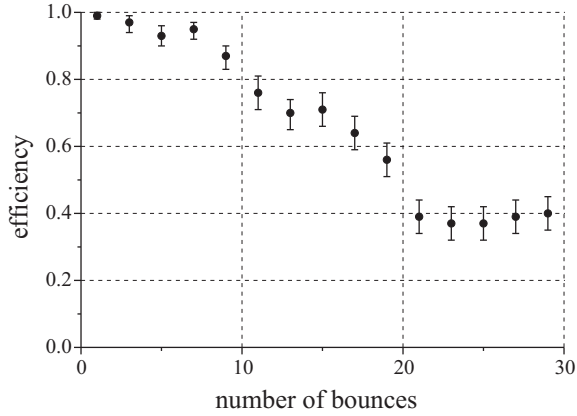}
\caption{'Single-atom shuttle'. The measured efficiency is plotted
versus the number of bounces, with a displacement distance of 1~mm
each and with an acceleration of 5000~m/s$^2$.}\label{figure6}
\end{figure}

More complicated position manipulations than a simple displacement
of an atom along the standing wave are possible. We implemented an
'atomic shuttle' by transporting one atom by 1~mm and then
reversing the direction of motion repeatedly. After swapping the
atom back and forth $n$ times, it is directly detected using
resonant illumination (Fig. \ref{figure6}). Each acceleration
process causes heating of the atom as discussed below which
results in a decreased detection efficiency. Heating processes are
sufficiently small, such that a single atom bounces back and forth
30 times with a measured efficiency of 40~\%.

\section{Acceleration}
\begin{figure}
 \centering
\includegraphics{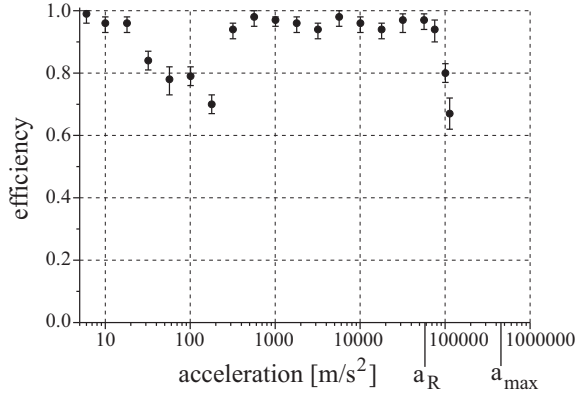}
\caption{Transportation efficiency for a displacement of 1~mm as a
function of the acceleration. The transportation efficiency for a
displacement of 1~mm remains well above 90~\% when the
acceleration is varied over several orders of
magnitude.}\label{figure7}
\end{figure}

We have investigated the transportation efficiency as a function
of the acceleration for a constant displacement of 1~mm using
resonant illumination detection. Although the acceleration $a$ was
varied over 4 orders of magnitude (Fig. \ref{figure7}), we found a
nearly constant transportation efficiency of more than 90~\% for
$a < 7\cdot10^4$~m/s$^2$. For larger accelerations, the efficiency
rapidly decreases.

The potential in the accelerated frame is the sum of the periodic
potential of the standing wave and the contribution of the
accelerating force, $U(z')=U_{0}\cos^2(kz')+maz'$. In the ideal
case of an initially motionless atom, the acceleration could be
adiabatically increased until the local minima of the standing
wave disappear. The acceleration is thus fundamentally limited by
the potential depth, $a_{\rm max}=U_0k/m=4.8\cdot10^5$~m/s$^2$.

\begin{figure}
 \centering
\includegraphics{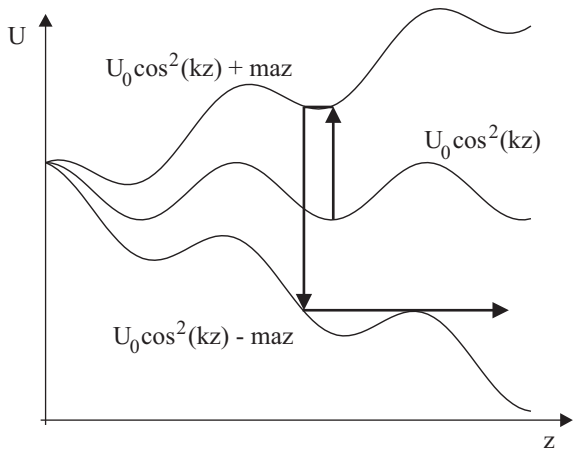}
\caption{Standing wave potential shown in reference frames
accelerated with $+a,0,-a$, respectively. Here the worst case of
maximal energy gain at $a=0.42\,a_{\rm max}$ is illustrated: Due
to the abrupt changes of the accelerating potential an initially
motionless atom can gain enough energy to leave the
trap.}\label{figure8}
\end{figure}

However, there are two effects that experimentally limit the
maximum acceleration to a lower value. The first is an additional
heating effect due to abrupt changes of the acceleration. At the
beginning of the transportation process the acceleration is
instantly switched from 0 to $a$ and similarly from $a$ to $-a$
after half the transportation distance (Fig. \ref{figure3}). The
equilibrium position of the accelerated potential is shifted by
the amount $\Delta z=-(2k)^{-1}\arcsin(a/a_{\rm max})$ (Fig.
\ref{figure8}). These sudden changes of the potential, which occur
three times during a transportation process, either heat or cool
the atom depending on the phase of its oscillation. The atom can
increase its energy twice during the first two changes, as
illustrated in Fig. \ref{figure8} where the worst case of a
maximal energy gain is shown. This leads to a non-zero probability
to lose the atom for accelerations exceeding $0.42 \,a_{\rm max}$.
The initial thermal energy of the atom, corresponding to
$0.15\,U_0$ in our case \cite{Alt 01b}, reduces this value to
$0.24\,a_{\rm max}$. For accelerations below $0.1\,a_{\rm max}$
this heating effect can be neglected. The maximum energy gained
due to abrupt jumps of the acceleration for $a\ll a_{\rm max}$ is
$\Delta E_{\rm max}(a)=4U_0(a/a_{\rm max})^2$ so that $\Delta
E_{\rm max}(0.1\,a_{\rm max})$ is only $80~\mu$K. Changing the
acceleration slowly enough would avoid this heating effect
completely because the atom would then adiabatically follow the
motion of the potential well.

The second fact that experimentally limits the maximum
acceleration is the finite bandwidth of the AOMs. The highest
acceleration achieved requires a mutual detuning of up to $20$~MHz
of the counter-propagating laser beams. For these parameters, the
AOM deflection efficiency decreases by 50~$\%$ which results in a
similar decrease of the dipole trap laser power and thus of the
trap depth. Due to the combination of these two effects the
transportation efficiency should decrease for accelerations
exceeding $0.12\,a_{\rm max}=5.8\cdot10^4$~m/s$^2$. This is indeed
what we observe. Note, however, that the accelerations realized
here still exceed that of the maximum resonant light pressure
force, $a_{\rm R}=p_{\rm ph} \Gamma/2m \approx 6 \cdot
10^4$~m/s$^2$. Here, $p_{\rm ph}=\hbar k_{\rm D2}$ is the photon
momentum. This allows us to change the atomic velocity from zero
to the maximum velocity of 10 m/s (limited by the AOM bandwidth)
within 100~$\mu$s.

The decrease of the transportation efficiency around 100~m/s$^2$
is attributed to a modulation of the dipole trap potential caused
by partial reflection of one of the laser beams on the glass cell.
The reflected beam interferes with the standing wave which is thus
phase and amplitude modulated. For appropriate detunings this
effect causes either resonant or parametric heating of the atoms.
This mechanism can be used to measure the oscillation frequency of
the atoms in the trap and is currently under study \cite{Alt 01b}.
If required, the effect could be avoided by slightly changing the
geometry of the setup or by selecting proper detunings of the two
beams.

\section{Conclusions and outlook}

We have realized a deterministic source of cold atoms which
delivers a prescribed number of single atoms to a desired spot.
The precise control of the experimental parameters allows us to
transport a single atom over the distance of 10~mm within 1~ms
with an accuracy of $\lambda/2$. The absolute position of the atom
is limited by the size of the MOT. The decrease in the efficiency
at larger distances is accredited only to the effect of gravity.
This indicates that larger displacements could be achieved by
using higher laser power and different beam geometries. Moreover,
additional cooling of the trapped atoms would further improve the
performance of the atomic conveyor-belt.

Alternative approaches are possible to control the motion of the
standing wave interference pattern. The first is to retro-reflect
one of the trap laser beams by a mirror. Translating the mirror
would result in a travelling wave. This mechanical solution would
avoid the major heating effect due to the phase noise of the AOM
drivers. Its disadvantages, however, are a worse overall stability
and much lower possible accelerations and velocities. A second
alternative is to use an EOM to phase-shift one of the laser
beams. Here, an adiabatic shift from 0 to $2\pi$ moves the
interference pattern along with the trapped atom by $\lambda/2$.
Then, the EOM rapidly switches the phase from $2\pi$ back to 0
without the atom being able to follow. Repeating this procedure
$n$ times transports the atom over the distance of $n\lambda/2$.
This approach would simplify the optical setup at the cost of
placing demanding requirements on the EOM driver.

One of the most interesting applications of the optical
conveyor-belt is the controlled positioning of two or more atoms
in the fundamental mode of a high-finesse optical cavity. In
current single-atom cavity-QED experiments \cite{Hood 00,Pinkse
00} atoms are released from a MOT and thus enter the cavity in a
random way.
%A string of atoms with well defined separations,
%created by the consecutive loading of single atoms into the dipole
%trap, could be a powerful tool in this context.
Our device, however, could place a predetermined number of atoms
into the cavity deterministically. This could provide the
possibility to entangle neutral atoms via the exchange of optical
photons \cite{Pellizzari 95,Beige 00}, a demanding task which, so
far, has only been accomplished in the microwave domain
\cite{Hagley 97,Rauschenbeutel 00}.

\begin{acknowledgement}
We have received support from the Deut\-sche
Forschungsgemeinschaft and the state of Nordrhein-West\-falen.
\end{acknowledgement}

\newpage

\onecolumn

\noindent{\bf Figures}

\twocolumn

\end{document}